\documentclass[aps,prl,twocolumn,groupedaddress]{revtex4}

\usepackage{amsmath, amsthm, amssymb}
\usepackage{amsfonts}
\usepackage{color}
\usepackage{graphicx}
\usepackage{makeidx}
\usepackage{url}

\newcounter{proposition}[section]
\newcounter{theorem}[section]
\newcounter{lemma}[section]
\newcounter{definition}[section]
\newcounter{remark}[section]
\newcounter{conjecture}[section]
\newcounter{corollary}[section]

\def\theproposition{\thesection.\arabic{proposition}}
\def\thetheorem{\thesection.\arabic{theorem}}
\def\thelemma{\thesection.\arabic{lemma}}
\def\thedefinition{\thesection.\arabic{definition}}
\def\theremark{\thesection.\arabic{remark}}
\def\theconjecture{\thesection.\arabic{conjecture}}
\def\thecorollary{\thesection.\arabic{corollary}}

\def\s #1 {\section{#1}}

\def\ssa #1 {\ifhmode{\par}\fi\refstepcounter{subsection}
  \noindent {\bf\thesubsection}. {\em #1}.\quad
  \addcontentsline{toc}{subsection}{\protect\numberline{\thesubsection} #1}%
  }

\def\ssb #1 {\ifhmode{\par}\fi\refstepcounter{subsection}
  \noindent {\bf\thesubsection.} {\em #1.}\quad
  \addcontentsline{toc}{subsection}{\protect\numberline{\thesubsection} #1}%
  }

\def\proposizione {\ifhmode{\par}\fi\refstepcounter{proposition}
  \noindent {\bf Proposition \theproposition}. \quad}
\def\teorema {\ifhmode{\par}\fi\refstepcounter{theorem}
  \noindent {\bf Theorem \thetheorem}. \quad}
\def\lemma {\ifhmode{\par}\fi\refstepcounter{lemma}
  \noindent {\bf Lemma \thelemma}. \quad}
\def\definizione {\ifhmode{\par}\fi\refstepcounter{definition}
  \noindent {\bf Definition \thedefinition}. \quad}
\def\remark {\ifhmode{\par}\fi\refstepcounter{remark}
  \noindent {\bf Remark \theremark}. \quad}
\def\congettura {\ifhmode{\par}\fi\refstepcounter{conjecture}
  \noindent {\bf Conjecture \theconjecture}. \quad}
\def\corollario {\ifhmode{\par}\fi\refstepcounter{corollary}
  \noindent {\bf Corollary \thecorollary}. \quad}

\begin{document}

\title{ Probabilistic Turing Machine and Landauer Limit }
\author{Marco Frasca}
\email[e-mail:]{marcofrasca@mclink.it}
\affiliation{Via Erasmo Gattamelata, 3 \\
             00176 Roma (Italy)}

%


\date{\today}

\begin{abstract}
We show the equivalence between a probabilistic Turing machine and the time evolution of a one-dimensional Ising model, the Glauber model in one dimension, equilibrium positions representing the results of computations of the Turing machine. This equivalence permits to map a physical system on a computational system providing in this way an evaluation of the entropy at the end of computation. The result agrees with Landauer limit.
\end{abstract}


\maketitle

\section{Introduction}

The idea that a classical computational system should always be strictly linked to a physical system was translated by Rolf Landauer onto a conjecture about its entropy. In 1961 he formulated a fundamental postulate \cite{land} that there exists a minimum entropy associated to a single bit of information. This quantity is $K\ln 2$ being $K$ the Boltzmann constant. This conjecture is fundamental as, otherwise, one should expect that, using information, the second principle of thermodynamics could be escaped. But, as pointed out by Bennett, Maxwell daemon fails just for this reason: The cost of information. Quite recently, it has been shown, using a smart experimental setup, that the Landuer conjecture is correct \cite{lutz}.

Due to this fundamental role Landauer conjecture has, our aim is to give a theoretical derivation of the limit he obtained on entropy. A possible way to give a theoretical foundation to this conjecture is to find a connection with a theoretical computational device: A Turing machine \cite{tur1,tur2}. A Turing machine is a deployable conceptual computational system that is able to perform the same computations as a real device can. Due to this completeness we need to map a physical system onto it and to show that the lower limit for entropy is the Landauer limit. So, our result will be founded on the Church-Turing thesis \cite{chur1,chur2}.

We will show that a probabilistic Turing machine can be defined that maps onto a one-dimensional Ising model: The Glauber model. In this way we can map dynamics and attach the meaning of physical quantities to a Turing machine.

\section{Turing voter}

As stated into the introduction, we base our conclusions on the Church-Turing thesis modified into the feasibility thesis that a probabilistic Turing machine can efficiently simulate any realistic model of computation. 

A probabilistic Turing machine (PTM) will be characterized by a random sequence of bit to be consulted to decide the next transition. We can write down it by a tuple $\langle Q, \Sigma, \iota, \sqcup, A, \delta \rangle$ for the Turing machine, where
\begin{itemize}
\item $Q$ is a finite set of states
\item $\Sigma$ is a finite set of symbols (the tape alphabet)
\item $\iota \in Q$ is the initial state
\item $\sqcup \in \Sigma$ is the blank symbol
\item $A \subseteq Q$ is the set of accepting states
\item $\delta \subseteq \left(Q \backslash A \times \Sigma\right) \times \left( Q \times \Sigma \times \{L,R\} \right)$ is a relation on states and symbols called the ``transition relation''.
\end{itemize}
To this machine we had a set of random symbols to be consulted for the next transition.

For our aims we will be interested in a machine with a single tape, two symbols $\Sigma=\{-1,1\}$ and a random transition mechanism that we will explain in a moment. Indeed, we assume the following randomness rule:
\begin{enumerate}
\item Select a random cell.
\item The selected cell assumes the symbol of the next-neighbor cell.
\item Repeat preceding steps without halting or when an acceptance or rejection is obtained. 
\end{enumerate}
This particular Turing machine can be called a ``voter'' as it simulates the behavior of a voting population \cite{katz}. So, the action of the machine can be to change the symbol as $1\leftrightarrow -1$ or to keep the symbol. This machine will be characterized by a rate of transition that can be written down as
\begin{equation}
   w(x)=\frac{1}{2}\left[1-x_i(x_{i+1}+x_{i-1})\right]
\end{equation}
and so the rate of flipping a symbol is $\frac{1}{2}$. This can also be generalized to
\begin{equation}
   w(x)=\frac{1}{2}\left[1-\frac{\gamma}{2}x_i(x_{i+1}+x_{i-1})\right]
\end{equation}
being $\gamma$ a factor expressing the tendency to change to the same symbol rather than the opposite depending on its sign. In this way we can introduce a master equation that determines the time evolution of the probability distribution for a given set of symbols of the machine when we start from an initial set $\{x^{(0)}\}$. We can write
\begin{equation}
   \frac{dP(\{x\})}{dt}=-\sum_iw(x_i)P(\{x\})+\sum_i w(-x_i)P(\{x\}_i)
\end{equation}
being $\{x\}_i$ the configuration $\{x\}$ but with i-th symbol changed. This is the dynamical equation for this probabilistic Turing machine. This machine describes the evolution in time of a population of voters that can reach a consensus determined by some halting condition. This kind of machine will be essential when we will discuss the physics of information.

\section{Non-equilibrium Ising model}

Ising model is a physical model representing the behavior of a ferromagnet. It is given by the following Hamiltonian
\begin{equation}
   H=-J\sum_{\langle ij\rangle}s_is_j-h\sum_is_i
\end{equation}
being $s_i$ spin variables taking the values $\pm 1$ depending on their orientation. The first sum is intended to run on the next-neighbors. The one dimensional case takes the simple form
\begin{equation}
   H=-J\sum_is_is_{i+1}-h\sum_is_i
\end{equation}
and is exactly solvable. This model has also a relaxation dynamics given by the Glauber model \cite{glau}. The master equation has an identical form of the voter master equation provided
\begin{equation}
\label{eq:map}
   \gamma = \tanh(2\beta J)
\end{equation}
being $\beta=\frac{1}{KT}$ with $T$ the temperature and $K$ the Boltzmann constant. On this basis we can state the following:
\vspace{0.3cm}

\begin{teorema}[Ising-Turing equivalence]
\label{teo1} 
A Turing voter is equivalent to a Glauber model.
\end{teorema}

\begin{proof}
A Turing voter, having an alphabet with two symbols, can always be mapped on a Glauber model as they evolve with the same master equation, provided eq.(\ref{eq:map}) holds, and symbols of the Turing voter have the same values the spin variables have.
\end{proof}

This theorem states that a computational machine like a Turing voter can always be mapped on a physical system, in this case a Glauber model representing the time evolution of a one-dimensional spin system. The equivalence of the end of computation is represented by any equilibrium state can be reached by the Glauber model in the course of its evolution. A one-dimensional Ising model has no critical point so, any possible configuration is a possible equilibrium and then also the end of a computation of the Turing machine.

\section{Landauer limit}

The relevance of the mapping provided in the theorem of the preceding section relies on the fact that we can map a physical system with a Turing machine. The probabilistic Turing machine can realize any computation a deterministic Turing machine can, the difference being just on complexity, and this is indeed the case of the Voter as, from any given starting configuration, a final configuration can always be obtained by the dynamic we have chosen. This imply in turn that we are in the condition to get a lower limit to the computational entropy needed for any computation, depending on the number of cells we consider. We can state the following:
\vspace{0.3cm}

\begin{teorema}[Landauer limit]
\label{teo2} 
The lower limit for the entropy of a Turing voter is the Landauer limit.
\end{teorema}

\begin{proof}
At the end of the computation, provided it worked at a temperature $T$, the Turing voter will have reached a configuration that is that of an Ising model with N spins at equilibrium at the same temperature. This model has a free energy at its equilibrium configuration
\begin{equation}
   F=-NkT\left[\ln 2+\frac{N-1}{N}\ln\cosh\left(\frac{J}{kT}\right)\right]
\end{equation}
and for the voter $N$ is just the number of cells. We get for entropy
\begin{eqnarray}
    S&=&-\frac{\partial F}{\partial T}= Nk\ln 2+(N-1)k\ln\cosh\left(\frac{J}{kT}\right) \nonumber \\
    &+&(N-1)\frac{J}{T}\tanh\left(\frac{J}{kT}\right) \nonumber \\
    &\ge& Nk\ln 2
\end{eqnarray}
and we recognize the minimum on entropy conjectured by Landauer to N bits of information. We note that the equality is attained for $N=1$.
\end{proof}
We note that the Landauer limit is independent on the Ising coupling $J$ and so, on the way we choose the parameter $\gamma$ into the mapping (\ref{eq:map}). We can always parametrize $\gamma$ through $\tanh(\beta J)$ and introduce in this way the dependence on the temperature for the Turing voter. Physically, this corresponds to the energy needed to the Turing machine to operate on the tape and a part of this energy will end up as heat. 

It is interesting to note that, the minimal energy one needs to erase $N$ bits at the temperature $T$ is just $NKT\ln 2$. At room temperature this is really a small number for known storage devices. So, for a petabit device one has about $10^{-6}\ J$ of minimum energy needed for erasing.

\section{Conclusions}

We have shown how Landauer limit can be obtained from a Turing machine, providing a mechanism to map a physical system, the Glauber model, onto the machine itself. In this way, the orignal conjecture by Landauer, experimentally proved quite recently \cite{lutz}, assume the nature of a mathematical theorem. Indeed, this is based on the idea that information is not an abstract entity but, when one is referring to real world, it must be just the dynamics of a physical system.

\section*{Acknowledgments}
I would like to thank Alfonso Farina for soliciting my interest on this area of research. Also, 
I would like to thank Joel David Hamkins for a really helpful answer to my question on this matter at MathOverflow \cite{MO98374}.

\end{document}